\documentclass[twoside]{IEEEtran}   %<---- Forces different odd/even pages, including first page
\usepackage[utf8x]{inputenc}
\usepackage{amsmath}

\ifCLASSINFOpdf
  \usepackage[pdftex]{graphicx}
\else
  \usepackage[dvips]{graphicx}
\fi

\usepackage{array}
\usepackage[printonlyused]{acronym}
\usepackage{cite}
\usepackage[tight]{subfigure}
\usepackage{multirow}
\usepackage{comment}
\usepackage{fancyhdr}

\usepackage{color, soul}
\usepackage{upgreek}
\title{Multi-Wideband Terahertz Communications\\ via Tunable Graphene-based Metasurfaces\\ in 6G Networks}

\author{Hamidreza~Taghvaee,~\IEEEmembership{Member,~IEEE,}
        Alexandros Pitilakis,
        Odysseas Tsilipakos,~\IEEEmembership{Senior Member,~IEEE,}\\
        Anna C. Tasolamprou,
        Nikolaos V. Kantartzis,~\IEEEmembership{Senior Member,~IEEE,}
        Maria Kafesaki,\\
        Albert~Cabellos-Aparicio,~\IEEEmembership{Senior Member,~IEEE}
        Eduard~Alarc\'{o}n,~\IEEEmembership{Senior Member,~IEEE,}
        and~Sergi~Abadal,~\IEEEmembership{Member,~IEEE,}
\thanks{H. Taghvaee is with the George Green Institute for Electromagnetics Research, Department of Electrical and Electronics Engineering, University of Nottingham, Nottingham NG7 2RD, United Kingdom (e-mail: hamidreza.taghvaee@nottingham.ac.uk)}
\thanks{A. Cabellos-Aparicio, E. Alarc\'{o}n and S. Abadal are with the NaNoNetworking Center in Catalonia (N3Cat), Universitat Polit\`{e}cnica de Catalunya, 08034 Barcelona, Spain}
\thanks{A. Pitilakis and N. V. Kantartzis are with the School of Electrical and Computer Engineering, Aristotle University of Thessaloniki, 54124 Thessaloniki, Greece, and with the Institute of Electronic Structure and Laser, Foundation for Research and Technology Hellas, 71110, Heraklion, Greece.}
\thanks{O. Tsilipakos, A. Tasolamprou and M. Kafesaki are with the Institute of Electronic Structure and Laser, Foundation for Research and Technology Hellas, 71110, Heraklion, Greece.}
\thanks{\textcolor{red}{\textcopyright \quad 2022 IEEE. Personal use of this material is permitted, but republication/redistribution requires IEEE permission. Refer to IEEE Copyright and
Publication Rights for more details.\\
Digital Object Identifier (DOI): 10.1109/MVT.2022.3155905}}

}% <-this % stops a space
\acrodef{RC}{Resistive-Capacitive}
\acrodef{CMOS}{Complementary Metal--Oxide--Semiconductor}
\acrodef{WNSN}{Wireless NanoSensor Network}
\acrodef{WSN}{Wireless Sensor Network}
\acrodef{MAC}{Medium Access Control}
\acrodef{QoS}{Quality of Service}
\acrodef{TS-OOK}{Time Spread On-Off Keying}
\acrodef{CSMA}{Carrier Sense Multiple Access}
\acrodef{TDMA}{Time Division Multiple Access}
\acrodef{FDMA}{Frequency Division Multiple Access}
\acrodef{CDMA}{Code Division Multiple Access}
\acrodef{ACK}{Acknowledgment message}
\acrodef{RF}{Radio-Frequency}
\acrodef{IR}{Impulse Radio}
\acrodef{OOK}{On-Off-Keying}
\acrodef{BER}{Bit Error Rate}
\acrodef{DVFS}{Dynamic Voltage and Frequency Scaling}
\acrodef{ARQ}{Automatic Repeat reQuest}
\acrodef{FEC}{Forward Error Correction}
\acrodef{M2M}{Machine-to-Machine}
\acrodef{mmWave}{millimeter-Wave}
\acrodef{THz}{Terahertz}

\begin{document}

%\pagestyle{fancy}
%\fancyhead[C]{\fontsize{7}{12} \selectfont \textcolor{red}{This work has been submitted to the IEEE for possible publication. Copyright may be transferred without notice, after which this version may no longer be accessible.}}

%\markboth{This work has been submitted to the IEEE for possible publication. Copyright may be transferred without notice, after which this version may no longer be accessible.}{\small  This work has been submitted to the IEEE for possible publication. Copyright may be transferred without notice, after which this version may no longer be accessible.}

\markboth{IEEE Vehicular Technology Magazine -- DOI:   10.1109/MVT.2022.3155905}{IEEE Vehicular Technology Magazine -- DOI: 10.1109/MVT.2022.3155905}

\maketitle

\begin{abstract} 
%Next-generation of wireless communication within the framework of 6G will be operational at the THz frequency regime. Although THz systems will dramatically enhance several performance indicators such as the data rate, spectral efficiency, and latency, exploiting such technology is challenging. In fact, Electromagnetic (EM) waves confront severe propagation losses like diffraction and atmospheric attenuation. Thus, relative communications are limited to Line-of-Sight (LoS) scenarios. In 5G networks, Reconfigurable Intelligent Surfaces (RISs) are introduced to solve this issue by redirecting the incident wave toward the receiver and implement Virtual-Line-of-Sight (VLoS) communications. In this paper, we aim to employ this paradigm for 6G networks and design a graphene-based RIS optimized to perform at multiple low atmospheric attenuation channels. Also, by tuning the biasing voltage of graphene, surface impedance is controlled and we can switch between the communication channels to isolate the links for specific applications. We further investigate the performance of this multi-wideband design through numerical and analytical analysis.  
The next generation of wireless networks is expected to tap into the terahertz (0.1--10 THz) band to satisfy the extreme latency and bandwidth density requirements of future applications. However, the development of systems in this band is challenging as THz waves confront severe spreading and penetration losses, as well as molecular absorption, which leads to strong line-of-sight requirements through highly directive antennas. Recently, reconfigurable intelligent surfaces (RISs) have been proposed to address issues derived from non-line-of-sight propagation, among other impairments, by redirecting the incident wave toward the receiver and implementing virtual-line-of-sight communications. However, the benefits provided by a RIS may be lost if the network operates at multiple bands. In this position paper, the suitability of the RIS paradigm in indoor THz scenarios for 6G is assessed grounded on the analysis of a tunable graphene-based RIS that can operate in multiple wideband transparency windows. A possible implementation of such a RIS is provided and numerically evaluated at 0.65/0.85/1.05 THz separately, demonstrating that beam steering and other relevant functionalities are realizable with excellent performance. Finally, the challenges associated with the design and fabrication of multi-wideband graphene-based RISs are discussed, paving the way to the concurrent control of multiple THz bands in the context of 6G networks.
\end{abstract}


\acresetall

%\begin{figure}
%  \centering
%  \includegraphics[width=\columnwidth]{../3.Figures/schematic.eps}
%  \vspace{-0.15cm}
%	\caption{Schematic diagram of a manycore processor integrating a wireline NoC and a WNoC.}
%  \label{fig:gwnoc}
%  \vspace{-0.5cm}
%\end{figure}

%\begin{table}[!t] 
%\caption{Manycore Communication Requirements}
%\caption{Wireless Manycore Scenario Requirements}
%\label{tab:reqs}
%\footnotesize
%\centering
%\begin{tabular}{ll} 
%\hline
%{\bf Metric} & {\bf Value} \\
%\hline
%Transmission Range & 0.1--10 cm \\
%Node Density & 10--1000 nodes/cm\textsuperscript{2} \\
%Network Throughput & 10--100 Gb/s \\
%Latency & 1--100 ns \\
%Bit Error Rate (BER) & 10\textsuperscript{-15} \\
%Energy & 1--10 pJ/bit \\
%\hline
%\end{tabular}
%\vspace{-0.3cm}
%\end{table}

%%%%%%%%%%%%%%%%%%%%%%%%%%%%%%%%%%%%%%%%%%

\section*{Introduction}
\label{sec:intro}

%%%%%%%%%%%%% New verison
Recent years have heralded the arrival of the fifth-generation (5G) of mobile and wireless communication systems with a plethora of technologies promising to address current data rate, latency, or density challenges. Yet, while 5G networks are still being deployed, the sixth generation (6G) is already under intense development aiming at continuing to satisfy the ever-growing communication demands of society and industry. In this context, it has been stated that 6G networks may increase the data rate by 10-50$\times$, the network efficiency by 10-100$\times$, the area traffic capacity by 100$\times$, while cutting down the latency by 10-100$\times$ with respect to that of 5G networks \cite{zhang20196g}. 

While 5G has tapped into the mmWave band (10--100 GHz) to combat the spectrum crunch, it is expected the 6G networks will need to enter the Terahertz band (0.1--1 THz) to continue increasing capacity and reducing latency through the exploitation of its ultra-broad bandwidth and the small form factor of its antennas. In this direction, THz-band communications will enable current networks to reach unprecedented locations, bridging them with nanonetworks within the human body or air-space networks composed of high-altitude platforms or satellite swarms \cite{zhang20196g}. This way, 6G is expected to play a key role in transportation (fully automated driving), immersive experiences (virtual and augmented reality, tactile internet, UHD streaming), or medicine (robotic surgery, eHealth) to cite a few examples.

\begin{figure*}[!ht]
    \centering
    \includegraphics[width=0.75\textwidth]{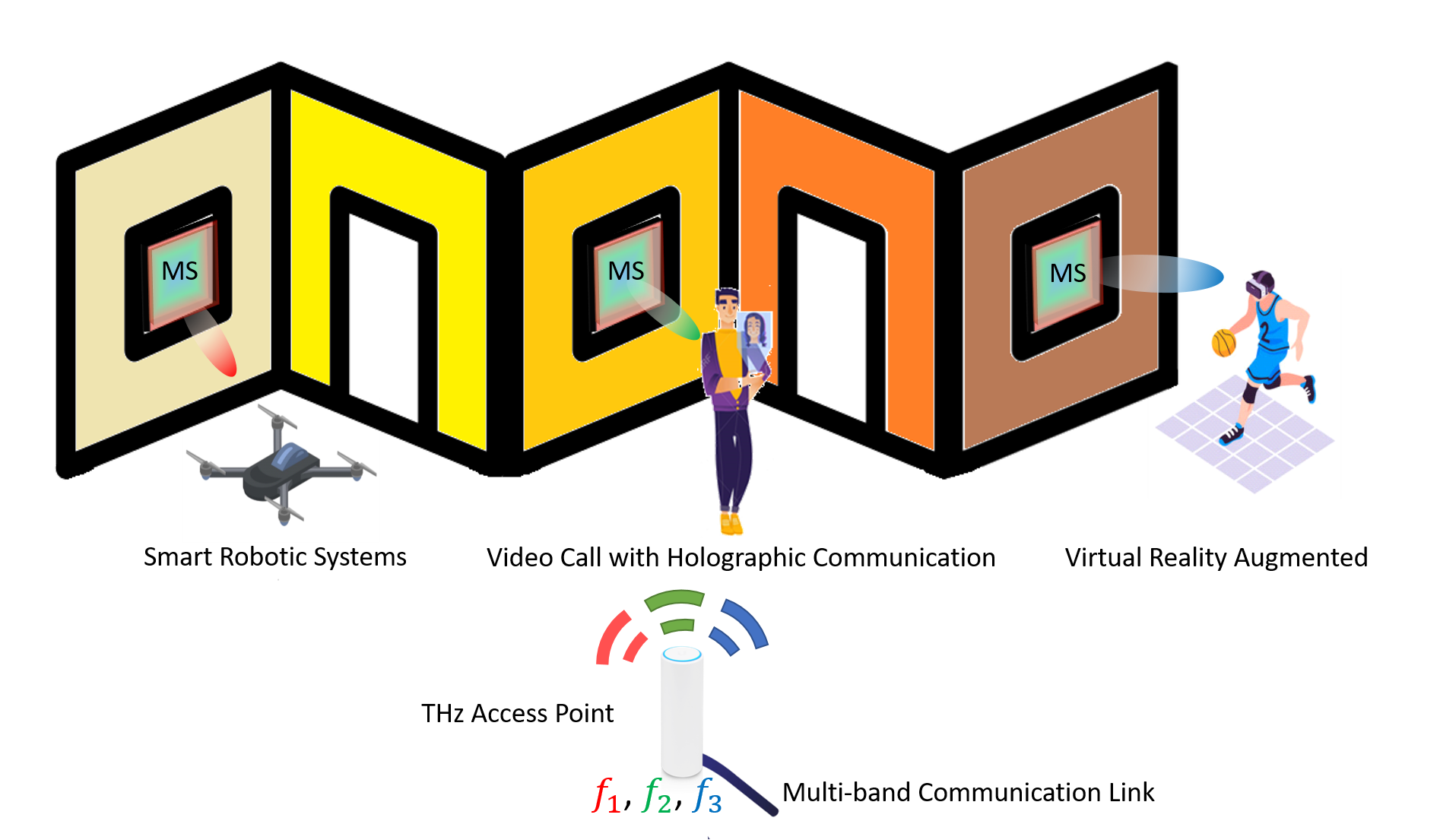}
    \vspace{-0.3cm}
    \caption{Graphical representation of an environment empowered with multi-wideband MS-based RIS, giving service to multiple users asking for divers applications requiring high data rate, ultra-low latency and reliable communications.}
    \label{fig:graphicalAbstract}
\end{figure*}

%We envision that 6G will not only enable a pervasively intelligent, scalable, reliable, and secure terrestrial wireless network but also can be incorporated for space communications such as satellite-to-satellite. 
In contrast to its multiple advantages, THz-band communications face important challenges. Among them, we highlight the very large attenuation suffered by propagating THz waves due to spreading losses, penetration losses, and molecular absorption. This imposes the use of directive antennas and strong Line-of-Sight (LoS) requirements that are hard to meet in dense scenarios. In this context, the concept of Reconfigurable Intelligent Surfaces (RISs) has been widely regarded as a promising solution to provide reliable, fast, and versatile communications at THz frequencies \cite{akyildiz2018combating}. To that end, a RIS is composed of an array of tunable unit cells whose collective response allows to control the strength, direction, and shape of the reflected waves. Such an array is referred to as a metasurface (MS) or a reflectarray depending on whether the separation between unit cells is much lower than the wavelength or not \cite{pan2021reconfigurable}.

Among the multiple uses of RIS in the context of THz-band networks, this paper focuses on the case of short-range and high-rate indoor communication as shown in Fig. \ref{fig:graphicalAbstract}. We envision a network in which a number of devices are connected through one or multiple non-contiguous wideband channels, most likely exploiting the transparency windows that exist in the THz band (see Fig. \ref{fig:att}), in a RIS-augmented environment. The use of bands may respond to aspects such as regulatory issues, gradual adoption of the THz band, the use of distance-aware modulations, or simply the need to maximize the network capacity using multiple carrier frequencies \cite{han2014distance}. In any case, the signals transmitted by the source, even with a highly directive pattern, will be subject to huge losses and therefore require the aid of the RIS to reach the receivers with enough power \cite{akyildiz2018combating}. The key observation here is that \textbf{\emph{the unique benefit provided by the RIS will be lost unless its internal MS/reflectarray is able to control the waves at the different frequency channels.} }
% and is located far from the users. To communicate with mobile users, this source usually uses a static or simple beamforming radiation pattern that distributes the Electromagnetic (EM) wave toward wide space. Then, received energy by the users is weak, and the established link is unstable for 6G standards. Nevertheless, by implementing multi-channel RISs with beamforming characters in the vicinity of the users, incident waves from the THz source can be redirected and more importantly increase the directivity. So, not only it can track the movement of the users in designated frequencies but also we can make sure the link is stable due to the directional radiation pattern.
%We posit that ongoing researches in the THz regime and reflective surfaces very well hold the key to the future of wireless communications.

The vision presented in this paper is that of an MS-based multi-wideband RIS, this is, a single device capable of controlling the amplitude, direction, and shape of the reflected beams \textbf{\emph{at multiple non-contiguous THz bands concurrently.}} To realize such a RIS, the constituting unit cells need to be widely tunable to provide the required amplitude/phase response at the different THz bands, for which tuning mechanisms used in lower frequencies are slow, bulky, or ineffective \cite{Qu2017}. Instead, graphene is an excellent candidate for this purpose thanks to its outstanding optoelectronic properties, which open the door to the development of ultrafast and highly integrated devices that naturally operate in the THz band and whose resonance can be tuned within a wide range by just adjusting a bias voltage \cite{wang2019graphene}. In fact graphene possesses distinct advantages with respect to common materials used for THz metasurfaces. Such metasurfaces may be composed of metallic, usually planar, elements with tunable response mediated by voltage controlled varying electronic interconnecting elements like diodes or properly selected layers infiltrated with voltage controlled liquid crystals. Compared to these configurations, graphene holds the advantage of being 
truly ultrathin while offering higher tuning speed and range. Another approach would involve the use of a  phase changing materials like vanadium dioxide which, with thermal stimuli, undergoes a change from insulator to metal. Despite the high 
tunability range this choice does not allow the required continuously changing local states and the very fast tunability of graphene. Tunable THz metasurfaces may also use photoconductive semiconductors  that offer high tuning speed, compared to that of graphene, but relatively low tuning range and would require a more complex system to control the local tuning.

% as a tunable 2D material for this purpose. On the surface of graphene, plasmons can propagate freely. The electrical conductivity can be tuned through chemical doping or electrostatic biasing. Plasmon tunability can be controlled precisely, in which case graphene-based MSs have been successfully integrated as reconfigurable designs for the manipulation of THz waves. 
%To realize such RISs, we need to design a multi-wideband MS capable of beam steering and collimation functionality for transmission and receiving the signal (TRx). In the transmission mode, incident wave impinging from the farfield is planar so with beam steering MS can redirect the wave toward the user. In the receiving mode, users can be assumed as point sources so with collimation MS can reform the beam front for the farfield THz source. 

In spite of their potential, work on graphene-based MS design has been mainly limited to a single functionality with a single-band operational frequency \cite{wang2019graphene, hosseininejad2019digital}. Here, in the pathway to demonstrating the proposed vision, we instead present and evaluate a RIS capable of operating at three distinct bands based on a widely tunable graphene-based MS. In a typical multi-carrier frequency system, there is  only one resonance in the frequency response spectrum of the hardware. There are some techniques for generating beams at very close frequencies but with limited bandwidth \cite{https://doi.org/10.1002/admt.202001032}. In contrast, we are able to tune the resonance frequency by a large amount while keeping the beamforming criteria for operation in the technologically uncharted THz band. The chosen bands are 0.56--0.74 THz, 0.75--0.98 THz, and 0.99--1.08 THz, identified in \cite{han2014distance} as feasible transparency windows (see Fig. \ref{fig:att}), which opens the door to high-speed transmissions to distances of several meters in all three bands, or even a few tens of meters in the lower band \cite{akyildiz2018combating}. In particular, we present a unit cell design able to provide the reflection amplitude and phase required to perform beam steering (to combat non-LoS propagation), beam splitting (for multicast transmissions), collimation (to implement RIS-based tunable antennas), or random scattering (to implement physical-layer security), which are later demonstrated. Hence, by supporting multiple bands and functionalities, a single RIS would be reusable and facilitate its adoption across different environments, protocols, and systems. However, the full realization of this vision poses multiple designs and implementation considerations, as we discuss in subsequent sections.

\section*{Graphene-based Multi-Wideband Metasurface}
%\section*{Graphene-based Widely Tunable Metasurface}
\label{sec:unit}

The behavior of an MS is the result of the collective response of its building blocks, namely, the unit cells. Since the proposed MS acts as a reflector, the unit cells need to yield a high reflection amplitude at all times. Further, for beam manipulation, we need to provide a unit cell with the ability to control the phase response over a wide range of values \cite{Qu2017}. Additionally, in our case, both conditions need to be met for multiple widely spaced frequencies, although not necessarily concurrently. To enable dynamic reconfigurability once the MS is deployed, it is necessary that both objectives can be met without physically changing the geometry. We thus propose that reconfigurability, both in frequency and phase, is achieved in the THz band by changing the biasing voltage of graphene sheets. 

\begin{figure}[!t] 
\centering
\vspace{-0.6cm}
\includegraphics[width=.8\columnwidth]{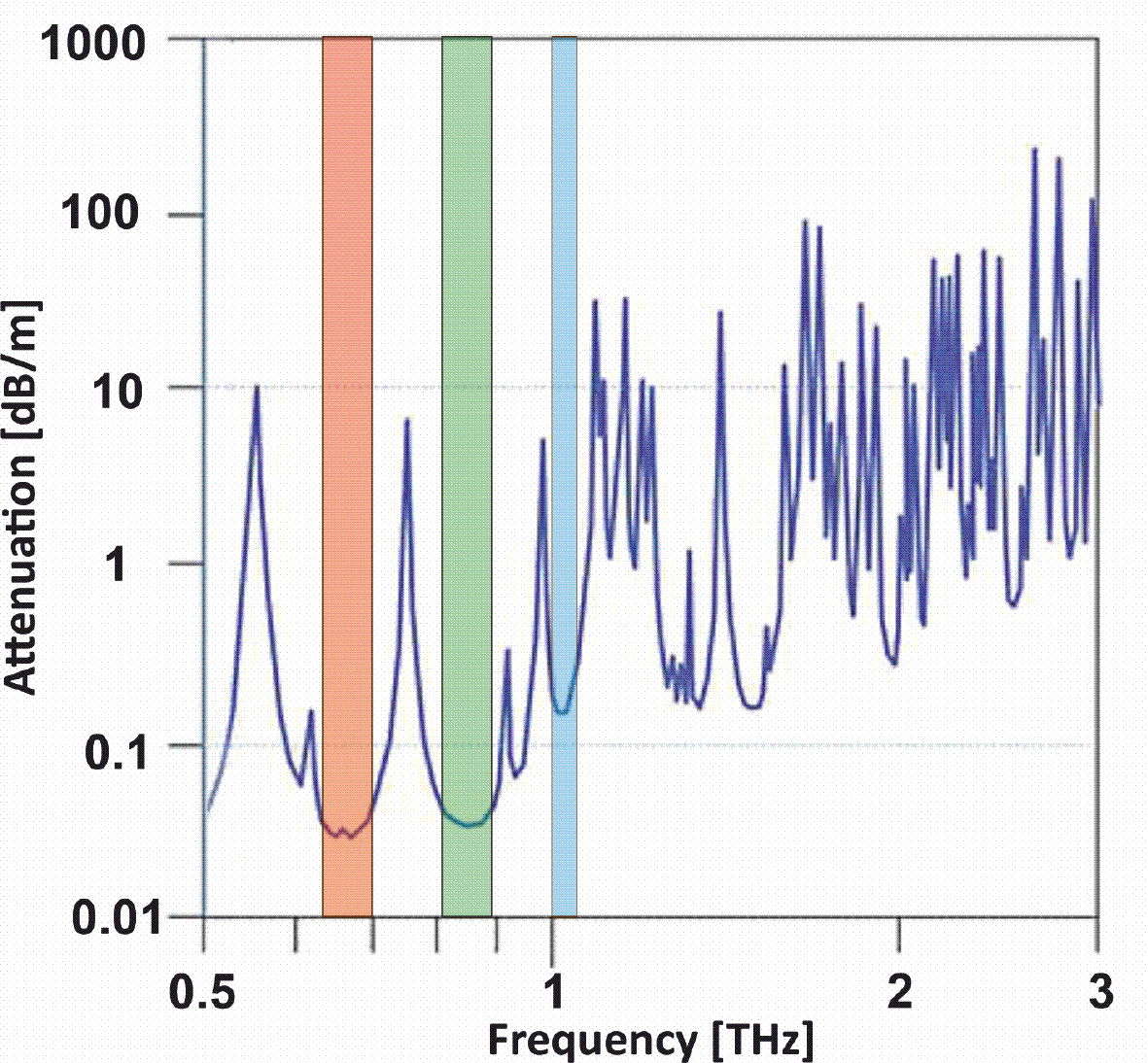}
\caption{Molecular absorption in a standard atmosphere at THz regime. Selected communication channels corresponding to the chosen transparency windows at 0.56--0.74 THz, 0.75--0.98 THz, and 0.99--1.08 THz, are highlighted in red, green, and blue. Source: HITRAN.}
\vspace{-0.3cm}
\label{fig:att}
\end{figure} 

Graphene exhibits a  broadband voltage-tunable metal-like surface conductivity thanks to its unique support of plasmonic modes in the THz band, which opens the door to the design of tunable MSs in a wide range of frequencies and phases by just changing a bias voltage \cite{wang2019graphene}.
%The unique plasmonic properties of graphene in the THz band allow to consideration of multiple spatial and frequency wireless channels. Moreover, SPPs in graphene are tunable electrically, which implies that the resonant frequency can be tuned within a wide range by just adjusting a bias voltage
It is worth noting that graphene's tunability margin begins to degrade in low THz frequencies. At the same time, due to the increasing free-space path loss at higher frequencies and the frequency-selective channel arising from the resonances of multiple elements present in a standard atmosphere, we cannot increase the operating frequency arbitrarily. Fig.~\ref{fig:att} shows the molecular absorption in a standard atmosphere in the THz spectrum, relative to the transmission distance. In light of this, the three highlighted windows at 0.56--0.74 THz, 0.75--0.98 THz, and 0.99--1.08 THz, are selected. In summary, to realize multi-band communication with an MS-based RIS, we need a unit cell that resonates at these frequencies and provides a wide reflection phase margin.

\subsection*{Unit Cell Design}
\begin{figure}[!t] 
\vspace{-0.4cm}
\centering
\includegraphics[width=0.5\textwidth]{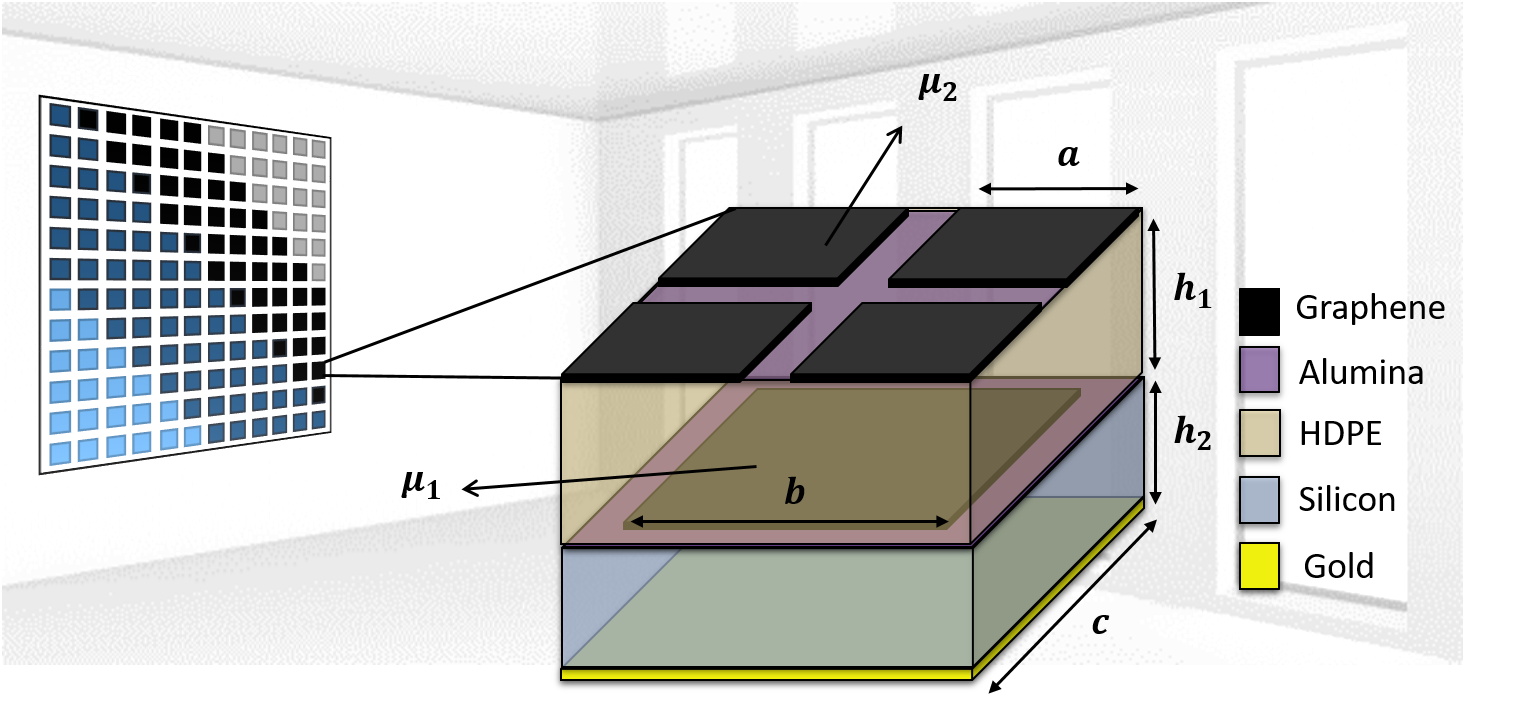}
\caption{Schematic representation of the unit cell layout. This unit cell is composed of two main dielectric layers (i.e., HDPE and silicon), four parasitic graphene patches on top and one graphene sheet sandwiched between the dielectrics. The unit cell is back plated with a thin layer of gold to be fully reflective.}
\label{fig:unit cells}
\vspace{-0.1cm}
\end{figure}

%%%It is worth noting that graphene's tunability margin begins to degrade in low THz frequencies. At the same time, due to the resonances of water molecules in THz frequencies, we need to be very careful about the selection of the operating frequency. Figure \ref{fig:att} shows the atmospheric attenuation propagation in the air at THz frequencies. It is apparent that high frequencies are more prone to propagation losses. In light of the above, three highlighted channels including 0.65, 0.85, and 1.05 THz are selected. To realize multi-band communication with MS-associated RIS, we need to design a unit cell that resonates at these frequencies and provides a wide reflection phase margin. %%\hl{SERGI: move this to the intro of the section.}
Fig.~\ref{fig:unit cells} shows the proposed unit cell whose lateral size is $c=35~\upmu$m (around $\lambda/10$), small enough to allow for fine resolution of the reflection phase profile, characteristic of MSs over standard reflectarrays. From top to bottom, the unit cell consists of a multi-layer structure with four parasitic graphene patches ($a=15~\upmu$m, chemical potential $\mu_2$) on top of an alumina layer with $0.1~\upmu$m thickness. Synthesis of graphene/alumina composite materials enhances strength, toughness, and wear-resistance with a low-cost process \cite{Kim2014}. This composite is stacked on high-density polyethylene (HDPE) substrate, due to its particularly low losses in the terahertz band \cite{polym12092094}. HDPE permittivity is $\epsilon_r=2.37$ and its thickness is selected $h_1=15~\upmu$m. The combination of graphene and HDPE leads to high electrical conductivity with good thermal and mechanical properties \cite{polym12092094}. Another layer of graphene/alumina composite ($b=20~\upmu$m, the chemical potential $\mu_1$) is sandwiched between HDPE and silicon ($h_2=10~\upmu$m) to allow for supporting multiple resonances and, thereby, being able to operate in three disjoint bands. 

Due to the native oxidation of silicon, a thin layer of silicon dioxide is grown in the outer layer in a short time. To analyze a realistic design we have considered a layer of silicon dioxide with $0.3~\upmu$m thickness on top of the silicon layer and the layout is back plated with gold. Regarding the fabrication feasibility of the proposed low-profile structure, there are advanced silicon substrate thinning techniques that can be used to achieve an ultra-thinning down to 4~$\upmu$m without damage occurred due to thinning processes. %\cite{kim2014ultra}. 

As mentioned above, the frequency and reflection phase of a graphene-based unit cell can be controlled via changes in its biasing voltage. This is modeled through the frequency-dependent surface conductivity of graphene $\sigma(f)$, which in the THz band is given by
\begin{equation}
\sigma\left(f\right)=\frac{e^{2}\mu}{\pi\hbar^2}\frac{i}{2\pi f+i\tau^{-1}},
\label{eq:sigma_graphene}
\end{equation}
where $e$ and $\hbar$ are constants, while $\mu$ and $\tau$ are variables that correspond to the chemical potential and the relaxation time of the graphene layer, respectively \cite{wang2019graphene}. When a graphene sheet forms one plate of a capacitor, any applied static voltage will alter the carrier density at its surface thus modulating its EM material properties, i.e., its complex-valued electric conductivity; the chemical potential, expressed in eV units, is typically used as the controllable variable, linked to the externally applied tuning voltage as $\mu\propto\sqrt{V_\mathrm{bias}}$. On the one hand, the amplitude response depends on the losses within graphene, $\mathrm{Re}\{\sigma\}$, which in turn depend on the quality of the graphene sheets as modeled by the relaxation time value $\tau$. For the purpose of this work, the relaxation time of graphene is assumed to be $\tau=1$~ps, which has been considered in multiple works and is realizable with state-of-the-art fabrication and encapsulation techniques \cite{Banszerus2015}. On the other hand, to modify the resonance frequency and the phase of the response, the key tuning variable is the chemical potential value $\mu$ or, in our case, two distinct values $\mu_{1}$ and $\mu_{2}$ on top and bottom patches as shown next. 
%The chemical potential can be controlled through electrostatic biasing using a DC voltage whose value depends on the surrounding materials and dimensions. 
All full-wave EM simulations at the unit-cell level were conducted in CST Microwave Studio (frequency domain solver).

%\begin{figure}[!ht] 
%\centering
%\subfigure[]{\includegraphics[width=0.49\columnwidth]{figures/A6.eps}
%\vspace{0.2cm}
%\includegraphics[width=0.49\columnwidth]{figures/P6.eps}
%\vspace{0.2cm}
%         \label{fig:a}}
%       \subfigure[]{\includegraphics[width=0.49\columnwidth]{figures/A8.eps}
%\includegraphics[width=0.49\columnwidth]{figures/P8.eps}
%         \label{fig:b}}
%  \subfigure[]{\includegraphics[width=0.49\columnwidth]{figures/A10.eps}
%\includegraphics[width=0.49\columnwidth]{figures/P10.eps}
%         \label{fig:c}}
%\caption{Amplitude and phase responses of reflection coefficient versus different chemical potentials of the sandwiched ($\mu_1$) and top ($\mu_2$) graphene layers for three selected operation frequencies such as 0.65 THz Fig. \ref{fig:a}, 0.85 THz Fig. Fig. \ref{fig:b}, and 1.05 THz Fig. \ref{fig:c}.}
%\label{fig:resultsUnit}
%\vspace{-0.1cm}
%\end{figure} 

\subsection*{Unit Cell States} \label{cell_states}
\begin{figure*}[!ht] 
\centering
\vspace{-0.5cm}
\includegraphics[width=0.65\textwidth]{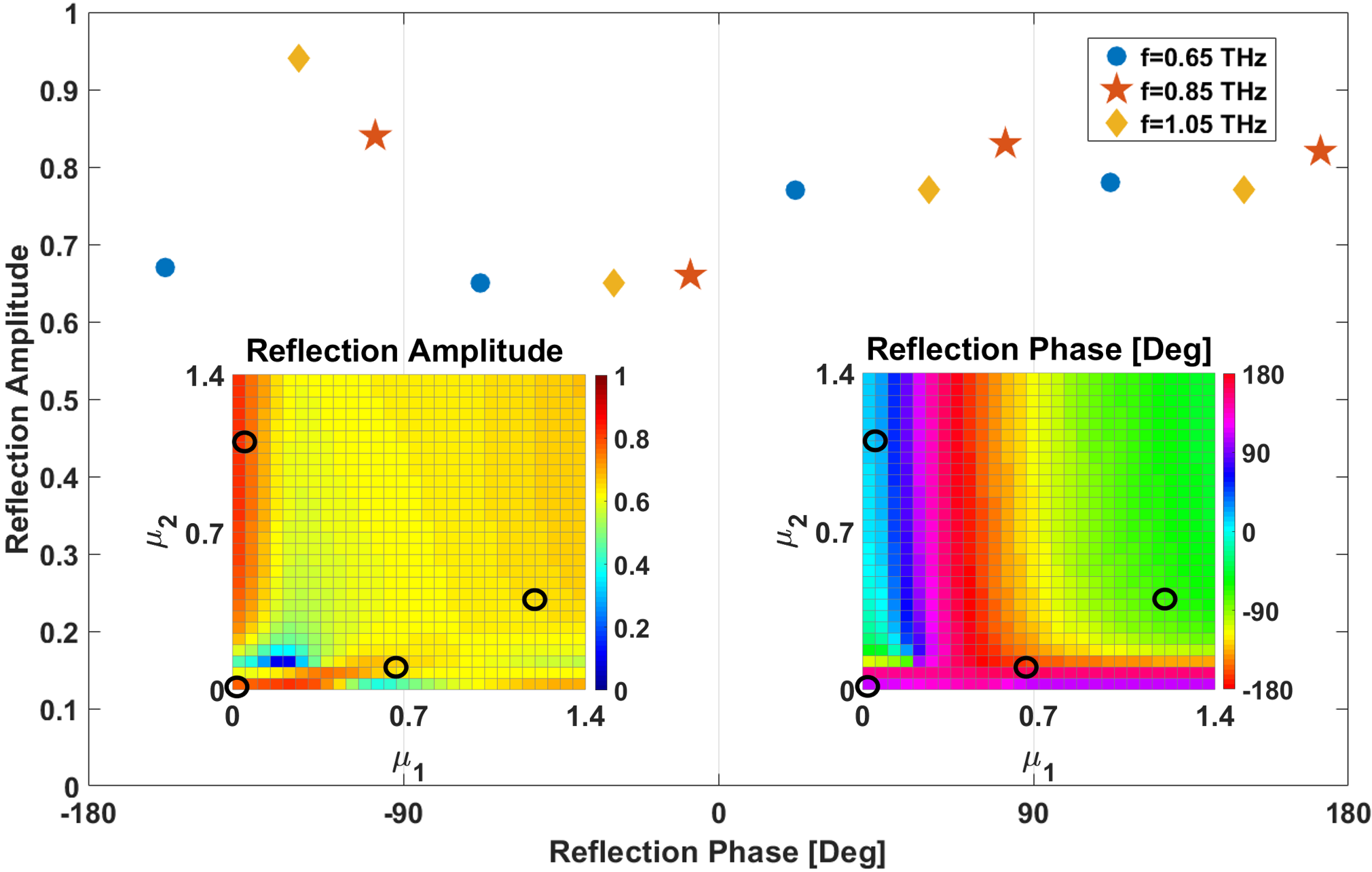}
\caption{Reflection characteristics of the selected states at respective operation frequency. The insets represent the design space of different combinations of chemical potentials at 0.65 THz. The horizontal axis represents the chemical potential of sandwiched graphene patch (see Fig. 3) and the vertical axis represents the chemical potential of the graphene patches on top. Four states corresponding to the highest possible amplitude and spaced with $\pi/2$ phase are selected (black circles in the inset and blue circles in the outset figure). Reflection amplitude and phase at 0.85 and 1.05 THz versus chemical potentials have similar design spaces. Selected states are plotted with red stars (0.85 THz) and yellow diamonds (1.05 THz) in the outset figure.}
\label{fig:apf}
\end{figure*}

We explore the chemical potential design space by sweeping the values of $\mu_{1}$ and $\mu_{2}$ in the three desired operational frequencies. The inset figures in Fig.~\ref{fig:apf} show a trade-off between reflection amplitude and reflection phase versus chemical potential values at 0.65 THz as an example. Hence, by controlling the bias voltage we can tune the reflection characteristics.

The results show the response of the MS for a continuous range of chemical potentials. However, to design a digital reconfigurable MS, we need to discretize the design space and select a set of chemical potentials to obtain a finite set of addressable states. Firstly, the number of unit cell states (per frequency band) has to be selected. Here, the number of states will determine the phase difference between consecutive unit cell states. As shown in our previous work \cite{9109701}, four states (with $\pi/2$ phase difference between them) can offer high-quality beam-steering performance, given that the unit cell is fairly subwavelength as in our case here. In the insets of Fig.~\ref{fig:apf}, black circles indicate the four selected states at 0.65 THz. These states are selected manually with two goals i) $\pi/2$ phase shift to each other and maximum reflection amplitude. The blue symbols in Fig.~\ref{fig:apf} represent the selected states at 0.65 THz, whereas the red and yellow symbols represent the other two frequencies.

We note that the selected states may need to change when the incidence angle changes. Assuming that a completely different set of states is needed to cover the neighborhoods of 0, 30, and 60 degrees \cite{9109701}, the minimum \emph{total} number of required states is 4 states per frequency band $\times$ 3 frequency bands $\times$ 3 incidence angle neighborhoods $=36$ states. Moreover, if the metasurface is brought in the near-field of transmitting/receiving antennas, the reflection phase/amplitude may vary and depend on the distance \cite{Danufane2021}. This behavior should be taken into account with additional states being foreseen to cover operation in such short distances.

\subsection*{Metasurface coding for beam steering} \label{sec:MScoding}
Having selected the four sets of chemical potentials $(\mu_1,\mu_2)$ that define the cell states, we retrieve the corresponding complex reflection coefficients (phase and amplitude) for each band, from charts similar to those in the insets of Fig.~\ref{fig:apf}, as calculated from full-wave EM simulations. In this work, we assume that the operation between the three bands is decoupled, i.e., the four $(\mu_1,\mu_2)$ sets that define the four states are different for each band. In this case, achieving independent control over the three bands is only possible via space multiplexing (large RIS panels divided into subpanels devoted to a particular frequency) or time multiplexing (slots reserved to each particular frequency) leveraging the fact that the response of graphene to the electrical tuning of its chemical potential is practically instantaneous. As discussed above, the phase-stepping is very close to $\pi/2$ along with the states of all three bands (horizontal axis of Fig.~\ref{fig:apf}), with a worst-case reflection coefficient of $-4$~dB; these metrics ensure qualitative agreement with theory and can be further improved with cell redesign.

The Huygens-Fresnel principle (HFP) \cite{9109701} allows us to evaluate the scattering pattern of the MS for a given configuration as the aggregated response of its unit cells to a given incidence (frequency, polarization, wavefront shape, and direction). For linearly polarized incidence waves, the scattered field can be expressed as
\begin{equation}
    \begin{split}
        E(\theta, \varphi) = \sum_{m=1}^{M} \sum_{n=1}^{N} A_{mn} p_{mn}(\theta_{mn}, \varphi_{mn})\\ R_{mn} p_{mn}(\theta, \varphi)
        e^{jk_0\zeta_{mn}(\theta, \varphi)}
    \end{split}
    \label{eq:Huygens}
\end{equation}
where $\varphi$ and $\theta$ are the azimuth and elevation angles, $A_{mn}$ corresponds to the complex amplitude of the wave incident for the $mn$-th unit cell, $p_{mn}$ denotes the scattering pattern of the $mn$-th unit cell, and $R_{mn}$ is the complex reflection coefficient for the $mn$-th unit cell. %which, according to reciprocity, is identical for scattering toward the $(\theta, \varphi)$ direction and the interception of incoming waves from the $(\theta_{mn}, \varphi_{mn})$ direction. 
Finally, $\zeta_{mn}(\theta, \varphi)$ is the relative phase shift of the unit cells with respect to the radiation pattern coordinates and their geometric arrangement on the MS. Tuning of the two voltages applied at each $mn$-cell alters graphene's local $\mu_{1,2}$ and consequently the cell's response, $R_{mn}$; the collective effect of the tuning of all $R_{mn}$ is a tuning of the scattering pattern, $E(\theta, \varphi)$. The underlying approximation of the HFP is that unit cells are uncoupled, so that their $R_{mn}$ can be tuned independently, which is valid for the cell sizes considered here. Finally, we note that HFP is an analytical method implemented on MATLAB using parameters calculated from 3D geometry and/or extracted from full-wave unit-cell level EM simulations.

The ideal MS coding or configuration, that is, the unit cell amplitudes and phases required to provide a given functionality, can also be analytically calculated in the same framework \cite{9109701}. This procedure can be summarized as forward-propagating the source `rays' while back-propagating the desired radiation pattern (prescribing a functionality) onto the MS, and then taking the complex-valued ratio of the propagated wavefronts on each cell; these ratios denote the required reflection coefficient $R_{mn}$ profile to realize the functionality.
Then, the unit cell states closest to the ideal ones are selected. As an example, normally incident plane waves can be steered away from the specular (normal) reflection direction by configuring the MS cells in stripes of progressively increasing reflection phase, according to diffraction grating theory. More elaborate beamforming patterns can be engineered by more complicated configuration patterns; for instance, beam-splitting in two different directions can be accomplished by superimposing the \emph{stripes} that correspond to single-beam steering thus generating a pattern of rectangular \emph{supercells}. 

\begin{figure}[!t] 
    \centering
    \vspace{-.1cm}
    \includegraphics[width=\columnwidth]{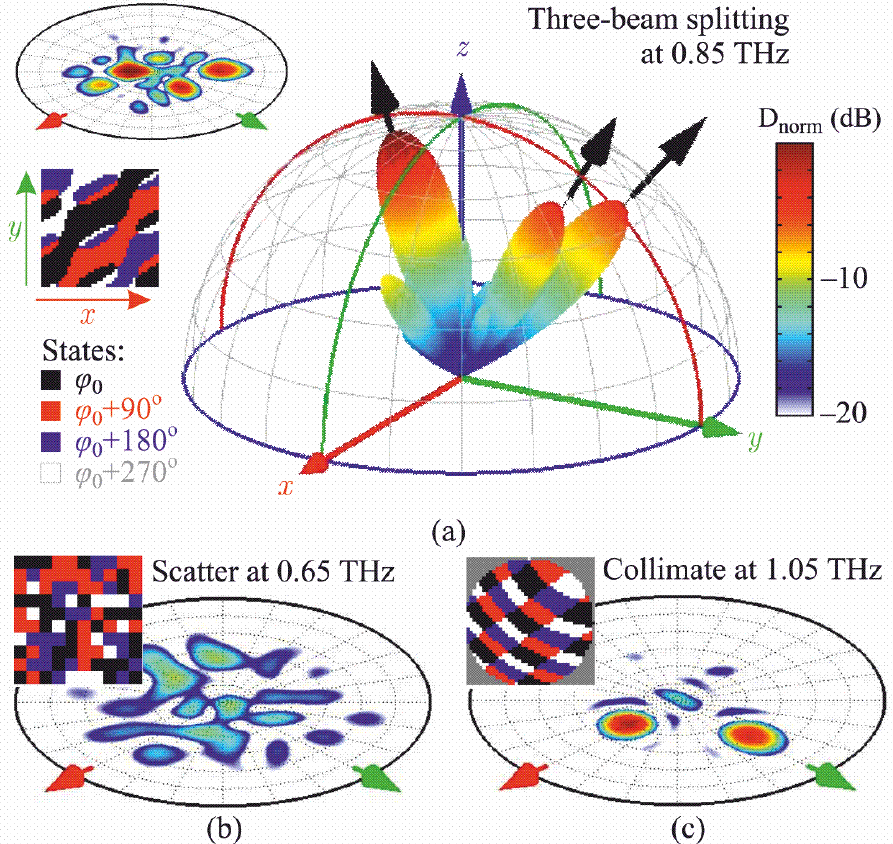}
    \caption{Scattering patterns for normal illumination of the multi-band multi-functional RIS; scattering heatmaps are normalized to theoretical maximum directivity. (a) 3D scattering pattern of a normally incident 0.85~THz plane-wave, split/steered in 3 predetermined directions: $(\theta,\varphi)=(15^\circ,315^\circ)$, $(30^\circ,90^\circ)$, and $(45^\circ,135^\circ)$; left-hand insets depict the cylindrical-coordinate representation of the same pattern, the orthogonal color-coded view of the MS cells configuration, and the color-mapping of the phase-states. (b) MS configuration and corresponding cylindrical scattering pattern for diffused scattering, and (c) MS configuration and corresponding cylindrical scattering pattern for collimation/steering of a spherical wavefront towards two distinct directions.}
    \label{fig:scatpats}
        \vspace{-.5cm}
\end{figure}

In Fig.~\ref{fig:scatpats}, we present a set of scattering patterns that demonstrate the potential of the proposed multi-wideband multi-functional RIS, assuming $40\times40$ 35~$\upmu$m-wide cells. This figure contains a selection of configurations and functionalities interesting in 6G contexts, demonstrated in three different frequencies, namely:
\begin{itemize}
    \item \textbf{Three-beam splitting} at 0.85 THz, Fig.~\ref{fig:scatpats}(a), shows a case where a normally incident plane wave is split in three beams, each one steered to a different direction that can be controlled independently of others: $(\theta,\varphi)=(15^\circ,315^\circ)$, $(30^\circ,90^\circ)$, and $(45^\circ,135^\circ)$. Fig.~\ref{fig:scatpats}(a) depicts the scattering pattern both in spherical (3D) and in cylindrical (2.5D) coordinates, in the main view and top-left inset, respectively, together with the unit cell states with color-coded squares in the left-hand insets. This functionality could be useful for multicasting in multi-band 6G networks. %the four-cell colors correspond to the complex reflection coefficients selected in Fig.~\ref{fig:apf} for the defined frequency, noting that they exhibit approximately a $\pi/2$ phase-stepping among the four states. 
%The response is similar to panel (a) but for 0.65 and 1.05~THz, presenting equally good performance and an increasing directivity (decreasing beam width) as the frequency increases, due to the fixed physical aperture of the MS whose electrical dimensions increase with frequency. 
    \item \textbf{Random scattering} of normally incident radiation at 0.65~THz, Fig.~\ref{fig:scatpats}(b). This is accomplished by totally random (uniform distribution) phase states across the MS; note that to eliminate back-scattering (or monostatic radar cross-section) the phase-states must be clustered in supercells of $\lambda/2$ size, as smaller randomly-coded cells emulate a metallic reflector and produce a specular reflection. This functionality is useful for physical-layer security methods, blocking certain users.
    \item \textbf{Collimation of a spherical wavefront towards two far-field directions} $(\theta,\varphi)=(30^\circ,0)$ and $(45^\circ,90^\circ)$ at 1.05~THz, Fig.~\ref{fig:scatpats}(c). In this case, the illuminating point-source is outside the near-field border, at 3~mm above the MS axis, so that emitted `rays' impinge almost perpendicularly on the MS. This functionality is useful for the building of antennas where an integrated RIS acts as an intelligent reflector.
    %Note also that we have emulated a spherical aperture by assigning full absorption (zero reflection amplitude coefficient) at the edges of the MS, as indicated by the grey-colored tiles, which has the advantage of reducing the side-lobes level. 
    \end{itemize}

%%%%%%%%%%%%%%%%%%%%%%%%%%%%%%%%%%%%%%%%%%

\section*{Implementation Considerations}
\label{sec:challenges}

%\subsection{Multi-band Coding Formulation}
The realization of the promising paradigm of widely tunable MS integrating graphene-based unit cells needs careful consideration of multiple aspects related to, among others, graphene technology and metasurface design. We discuss them next.

\subsection*{Technological implementation}
Although there are a rising number of methods for making various forms of graphene, the volume production for some of those methods remains limited. The focus will have to be on material quality and manufacturing cost. The main challenge lies in maintaining the quality of graphene, i.e. large $\tau$, in large areas. Defects on the graphene layer will significantly affect the electrical conductivity. %\cite{Izzaty_2019,,Lee2019ReviewOG}.
%Another factor that affects the electrical conductivity is the thickness of graphene. Misra et al. investigated the thickness dependency of the graphene features using  multiple monoatomic sheets as a metal gate electrode in a metal oxide semiconductor structure by inserting them between the dielectric (SiO2) and contact metal (TiN) \cite{doi:10.1063/1.4726284}.
Therefore, choosing the most accurate method for producing graphene and good control over the graphene layer deposition is very important. % \cite{C7CS00363C}.
Various fabrication approaches have been utilized to develop graphene. The more efficient among them, Chemical Vapor Deposition (CVD), has been shown to provide high quality, single layer, uniform graphene in large scale samples~\cite{Choi2017116X}. %while other approaches may involve micromechanical exfoliation, liquid-phase exfoliation, and pulsed laser deposition

The contact of graphene with other materials, for example, dielectric substrates or metallic electrodes, modifies the doping level and chemical potential of graphene. In this case, encapsulating graphene in insulating materials such as hexagonal Boron Nitride (hBN) helps minimizing the influence of contacting materials and, hence, maintaining a high $\tau$ \cite{wang2019graphene}. % \cite{PhysRevB.87.075414}.
%In this regard, Zhu et al. reported CNT films as a perfect substrate for the graphene layer to preserve its intrinsic work function \cite{Zhu2014}.
For the implementation of graphene-based MSs, it is usual to deploy a patterned scheme applied in a uniform sheet, as the patches discussed here. For the patterning, post-process techniques in pre-synthesized graphene are available, i.e., electron-beam, helium ion beam, lithography, nanostructure-assisted, or nanoimprint technology~\cite{Wei20201655}.

%\cite{Zheng2017,Bell2009,Liu20111246,Liang20102454}
%\hl{SERGI: We may need to tone this section down a bit as it is far from the audience. Anna: sure, i will rephrase}

\subsection*{Graphene tuning mechanisms}
 %Graphene exhibits exquisite electrical and optical properties; these stem from the linear dispersion of its 2D Dirac fermions, which allows for wide tunability. 
 As mentioned before, the properties of graphene depend strongly on aspects such as relaxation time and chemical potential, the latter of which can be tuned by imposing an external stimulus. Thus one can modify graphene's conductivity by altering the chemical doping, by thermal tuning, external electrostatic or magnetic field, and optical pumping or self-action. Many of the works investigating the physical aspects of graphene's modulation capabilities focus on the global modification usually involving uniform graphene sheets. The most prevalent technique for locally manipulating the optical properties of graphene is controlling the biasing voltage, often enhanced with the use of an electrolytic medium. In this case, care must be taken in controlling the biasing voltage at each unit cell individually as required by the MS coding. This non-uniform modification can be achieved by properly placed transparent electrodes or by alternative techniques, for example, uneven ground planes or homogeneous dielectric spacers~\cite{DeAbajo2014133}, which in any case are non-trivial at the granularity of the proposed unit cells. In all cases, we note that the tuning speed of graphene (in the order of picoseconds) is enough to support not only slow-paced indoor applications, but also the most demanding outdoor applications, even vehicular ones, whose reconfiguration speeds may be in the the order of hundreds of microseconds \cite{akyildiz2018combating}.
 %~\cite{Liu20113335,Sun2012114,Das2008210,Yu20093430,Tasolamprou2019720}. 
 
 %To this end, it is necessary to contact the graphene
%with other materials. In theory, the Fermi level of pristine graphene is considered to coincide with the Dirac point, while in real scenarios, graphene contact with a metal (or insulator) alters its electrical structure. Therefore, a full understanding of the physics of the metal (insulator)/graphene interfaces is very important \cite{PhysRevB.87.075414}. In this regard, Zhu et al. reported CNT film as a perfect substrate for the graphene layer for preserving its intrinsic work function \cite{Zhu2014}.

%\subsection{}

 %In theory, the Fermi level of pristine graphene is considered to coincide with the Dirac point, in this regard, Zhu et al. reported CNT film as a perfect substrate for the graphene layer for preserving its intrinsic work function \cite{Zhu2014}.

\subsection*{Multi-band metasurface design}
%\hl{SERGI: we could add things here if we just prove uncoupled operation. Here we could explain the challenges associated to coupled operation and how the unit cell design/biasing/states/coding would need to change to accommodate the three bands concurrently.}
%%\hl{ALEX to all: Is this OK?}

Assuming that the MS operation in all three bands is concurrent, e.g., without any time-multiplexing scheme to decouple them, means that setting the graphene chemical potentials $(\mu_1,\mu_2)$ to optimize the four states for one frequency band will unavoidably simultaneously affect the graphene response in the other bands. In line with the vision outlined in this paper, an ideal unit cell would provide a large number of states with high amplitude and all possible combinations of equally spaced phases in the three target frequencies, i.e. $\varphi=\varphi_{0}+\{0,\tfrac{\pi}{2},\pi,\tfrac{3\pi}{2}\}$ at $f_{1},f_{2},f_{3}$. This would allow controlling the response at the three bands simultaneously and independently. However, such a unit cell would probably require heavy optimization to strike a delicate balance among multiple resonances.

\begin{figure}[!t] 
    \centering
    \includegraphics[width=\columnwidth]{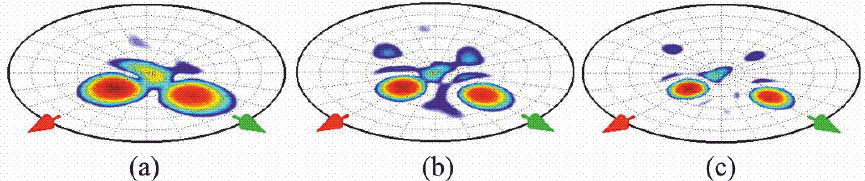}
    \caption{Scattering patterns for two-beam splitting of a normally incident plane wave by a circular-aperture MS ($\{a,b,c\}=\{0.65,0.85,1.05\}$~THz).}
    \label{fig:coupled_response}
\end{figure}

In less ideal unit cells, the complex reflection coefficients in the other bands will in principle form a sub-optimal set of states deteriorating the performance. It can be intuitively understood that optimization of the unit-cell geometry can potentially lead to adequately broadband behavior; alternatively, the four states can be assigned to four $(\mu_1,\mu_2)$ sets that lead to a tolerable performance, \textit{on average} across the operation bands. In Fig.~\ref{fig:coupled_response}, we present two-beam splitting of a normally incident plane wave by a circular-aperture MS (diameter 1.4~mm) of 35~$\upmu$m-wide cells. Each panel in this figure corresponds to an operating frequency band ($\{a,b,c\}=\{0.65,0.85,1.05\}$~THz) for which the $(\mu_1,\mu_2)$ for the four states have been optimally selected. Evidently, the scattering patterns present two well-defined lobes at the desired directions, $(\theta,\varphi)=(30^\circ,0)$ and $(45^\circ,90^\circ)$. In each case, the response at the bands for which the MS has not been tuned, which is not shown for space constraints, is degraded. Specifically, the scattering directivity in the desired lobes was decreased by 1-2~dB (worst-case) whereas parasitic side lobes were raised by 6-7~dB.
%\hl{Sergi's question: was the degradation a lot? Can we quantify it? Since we lost the charts due to space limitations, this part of the text is a bit lacking.}\textcolor{blue}{[See comment-Alex]}

%\textcolor{blue}{\hl{[Alex 22-Dec]}: I did some tests on the case of Fig. 6(b), 0.85 THz two-beam splitting. Below, I list the max-Directivity and SLL for these four cases: (i) Ideal states [90-deg phase-stepping and amplitude=1], (ii)-(iv) `real' states chosen optimally [from Fig. 4] for frequency 0.85, 0.65, 1.05 THz.}

%\textcolor{blue}{\textbf{max-Directivity dB}: (i) 19.3, (ii) 18.5, (iii) 17.1, (iv) 17.2}

%\textcolor{blue}{\textbf{SLL dB}: (i) 17, (ii) 13, (iii) 8, (iv) 7}

%\textcolor{blue}{The worst-case degradation w.r.t. to reference case (ii) [not the ideal case (i)] is 1.3~dB for max-D [negligible] and 6~dB for SLL. I should note that in case (ii) the undesired 13dB lobe was `back' towards specular direction [normal] negligible, whereas in cases (iii)-(iv) it was in a `mirror' direction to one of the two desired lobes. So, I think we can safely add the following text:}

\subsection*{Near-field and link-budget considerations}
% \hl{Which is the near-field distance and Fresnel distance? How does that compare to pico-cells or indoor scenarios? That would enable or actually make important the use of collimation and focusing.}

% \hl{ALEX: I have added some details for that in II.D, second paragraph (estimation of near-field and far-field distances). I can elaborate a bit on pico/femto-cells or indoor channels, or even use of this RIS as a part of an antenna (reflect-array), but I don't think it merits a subsection of its own, or I can't handle it. Maybe merge some sentences in the Introduction?}
% \textcolor{red}{Hamidreza to Alex: I think, Section II.D is longer than the rest of subsections. So, maybe it is a good idea to use the content about the nearfield in this subsection.}
%%\hl{ALEX to all: I merged the last two subsection in this, cropping some things from II.D}

The Huygens-Fresnel principle introduced previously can be used to accurately estimate the far-field as a superposition of elementary scattering patterns from each cell. %, assuming that their reflection coefficients are independent and the illuminating source is not in the near field. In this work, we assumed a $40\times40$ cell square MS of 35~$\mu$m-wide square cells, leading to a full-aperture diameter of $D=\sqrt{2}\times1.4$~mm. In this case, the near- and far-field limits are approximately at 3~mm and 25~mm, respectively, for 1~THz radiation. These distances are given by classic antenna-theory formulas, $r_{NF}=0.62\sqrt{D^3/\lambda}$ and $r_{FF}=2D^2/\lambda$, respectively. 
Even though this Huygens-Fresnel framework is strictly valid in the far-field, i.e., in a context similar to geometric-optics/ray-tracing, its applicability for performance evaluation extends well into the `transition' (or Fresnel) region, bounded by the near-field and the Fraunhofer ($r_{FF}$) regions. This means that a RIS can be envisioned as a part of the antenna system of a transceiver base station, providing digital beamforming that enhances path loss to virtual-line-of-sight mobile stations. Moreover, in real-world applications relevant to 6G THz communications, such as pico/femto-cells or short-range indoor channels, all incoming sources will practically be in the far-field of the MS, even for cm-sized apertures for which the Fraunhofer region is a few meters.

Finally, note that this multi-band RIS is capable of operating at frequency bands of almost an octave distance. In this sense, the path loss will exhibit a 6~dB deterioration switching from the lowest to the highest frequency, which could be considered for a short-range link budget. Nevertheless, this difference can be partially compensated by an increased directivity of the scattering pattern assuming that: (i) the RIS aperture is fixed, (ii) the channel is appreciably unobstructed so that a low exponent $\lambda^n$, $n\rightarrow2$, can be assumed, and (iii) the MS reflection coefficients are approximately similar across the frequency bands; the latter can be attained by engineering the unit-cell and careful selection of its states, and/or using materials with sufficiently broadband response, such as graphene.

\section*{Conclusions}
\label{sec:conc}

The concept of a multi-wideband MS in the THz regime has been explored in an attempt to extend the RIS paradigm to 6G networks operating at multiple frequencies concurrently. Embedding multiple layers of graphene into the design of the unit cells allows reconfiguring the local reflection phase at multiple frequencies by tuning the biasing voltage of each layer independently. With the aggregated response of the unit cells within the MS, we show that beam steering, beam splitting, random scattering, and collimation functionalities can be achieved at the desired frequencies (i.e., 0.65, 0.85, and 1.05 THz), which are characterized by low atmospheric attenuation. The realization of the proposed multi-band vision, however, still requires unit cells allowing for the concurrent yet independent control of the phases at each target frequency. Even though this complicates the design and implementation of the unit cells, the potential impact in the deployment and subsequent adoption of 6G networks is deemed considerable since a single RIS could be reused across environments, protocols, and systems.

%\section*{Acknowledgment}
%This work has been supported in part by the European Commission under grant H2020-FETOPEN-736876  (VISORSURF).

%\textcolor{red}{Refs List: Add Danufane2021 \cite{Danufane2021} asked by Reviewer3? Maybe it can replace some of the materials-related (graphene) references.}
%\hl{Odysseas - I have added it as Ref. [13] in the new discussion of the unit cell states and the dependence on angle of incidence. Since the editor is ok with the extra reference, probably there is no need to delete any reference.}

\bibliographystyle{IEEEtran}

\end{document}